\documentclass[useAMS,usenatbib]{mn2e}
\usepackage{graphicx}
\usepackage{lscape}
\usepackage{longtable}
\usepackage{rotating}
\usepackage{colordvi}
\usepackage{color}
\usepackage[english]{babel}
\usepackage{amsmath}
\usepackage{amssymb}
\usepackage{amsfonts}
\usepackage{epsfig}
\usepackage{subfig}
\usepackage{natbib}

\title[Testing $f(R)$-theories using the first time derivative of the orbital period of the binary pulsars]{Testing $f(R)$-theories using the first time derivative of the orbital period of the binary pulsars}
\author[M. De Laurentis, I. De Martino]{Mariafelicia De Laurentis $^{1,2}$\thanks{e-mail address: felicia@na.infn.it},
Ivan De Martino$^{1,2,3}$\\
$^{1}$ Dipartimento di Scienze Fisiche, Universit\`a di Napoli "Federico II"\\
$^{2}$ INFN sez. di Napoli Compl. Univ. di Monte S. Angelo, Edificio G, Via Cinthia, I-80126 - Napoli, Italy\\
$^{3}$\it Departamento de  Fisica Teorica, Universidad de Salamanca, 37008 Salamanca, Spain}
\begin{document}
\date{Accepted xxxx Yyyyber zz. Received xxxx Yyyymber zz; in original form xxxx Yyyyber zz}
\pubyear{2012}

\maketitle
\label{firstpage}

\begin{abstract}
In this paper  we use one of the Post-Keplerian parameters to obtain constraints on  $f(R)$-theories of gravity. Using Minkowskian limit, we compute the prediction of $f(R)$-theories on the first time derivative of the orbital period of a sample of binary stars, and we use our theoretical results to perform a comparison with the observed one. Selecting a sample of relativistic binary systems we estimate the parameters of an analytic $f(R)$-gravity. We find that the theory is not ruled out if we consider only the double neutron star systems, and in this case we can cover the  existing gap between the General Relativity prediction and the observed data.
\end{abstract}


\section{Introduction}
The gravitational waves (GWs) are one of  the most promising tools to study astrophysical systems like Neutron Stars (NS), 
coalescing binary systems, Black Holes (BHs), and White Dwarfs (WDs). 

The observational indirect evidences of gravitational radiation  were measured on the  system B1913+16, known as the Hulse-Taylor binary pulsar,  
and confirmed in others relativistic binary systems. The prediction of General Relativity (GR) on the first time derivative of the orbital period 
in binary pulsar systems was studied by \cite{ht75a} and \cite{wnt10}, for which the discrepancy on observed data with respect to the prediction is $\sim1\%$. However, the observational results should be explained using 
a different formulation of gravity \citep{fvf+12}. As shown in \cite{quadrupolo}, these systems could represent a good test for  
Extended Theories of Gravity (ETG).  Considering a class of analytic $f(R)$-theories, it is possible evaluate the gravitational radiated power 
in weak field limit. In this approximation we find that the energy radiated depends on the third derivative of the quadrupole, as predicted by GR, and the fourth derivative 
representing the corrective contribution to the theory. 
This result can be used to set constraints on the theory,  comparing the prediction on the first time derivatives of the orbital 
period with respect to the observed one. 
The outline of the paper is the following: in sec. 2 we briefly introduce the weak field limit approximation of $f(R)$-theories of gravity.
In sec. 3 we apply the theoretical results previously obtained  to binary systems computing the energy lost through GWs emission. 
In sec. 4 we compute the first time derivative of  the orbital period in $f(R)$-theories of gravity, and we compare the theoretical prediction with the observed data.
Finally, in sec. 5 we give our conclusions and remarks.

\section{f(R)-gravity background}
The $f(R)$-theories are based on corrections and enlargements of the GR theory adding higher-order curvature invariants and minimally 
or non-minimally coupled scalar fields into dynamics which come out from the effective action of quantum gravity 
 \citep{PRnostro}.

Starting from the following field  equations in $f(R)$-gravity (looking for major
details at \cite{PRnostro}, \cite{PRsergei}, \cite{nojiodi}, \cite{francaviglia}, \cite{faraoni})\footnote{${\displaystyle T_{\mu\nu}=\frac{-2}{\sqrt{-g}}\frac{\delta(\sqrt{-g}\mathcal{L}_m)}{\delta
g^{\mu\nu}}}$ is  the energy momentum tensor of matter ($T$ is the
trace), ${\displaystyle \mathcal{X}=\frac{16\pi G}{c^4}}$ is the coupling, ${\displaystyle f'(R)=\frac{df(R)}{dR}}$,
$\Box_g={{}_{;\sigma}}^{;\sigma}$, and $\Box={{}_{,\sigma}}^{,\sigma}$. We adopt a $(+,-,-,-)$
signature, and indicate with "$,$" partial derivative and with " $;$" covariant derivative with regard to  $g_{\mu\nu}$; 
all Greek indices run from $0,...,3$ and Latin indices run from $1,...,3$; $g$ is the determinant.}:
\begin{eqnarray}\label{fe1}
f'(R)R_{\mu\nu}-\frac{f(R)}{2}\,g_{\mu\nu}-f'(R)_{;\mu\nu}+g_{\mu\nu}\Box_g f'(R)\,=\,\frac{\mathcal{X}}{2}T_{\mu\nu}\,,
\end{eqnarray}
\begin{eqnarray}\label{TrHOEQ}
3\Box f'(R)+f'(R)R-2f(R)\,=\,\frac{\mathcal{X}}{2}T\,,
\end{eqnarray}
the Minkowskian limit can be calculate for a class of analytic $f(R)$-Lagrangian ({\it i.e.} Taylor expandable in term of
the Ricci scalar\footnote{For convenience we will use $f$ instead of $f(R)$. All 
considerations are developed here in metric formalism. From now on we assume physical  units $G=c= 1$.})
\begin{eqnarray}\label{sertay}
f(R)=\sum_{n}\frac{f^n(R_0)}{n!}(R-R_0)^n\simeq
f_0+f'_0R+\frac{f''_0}{2}R^2+...\,.
\end{eqnarray}
At the first order, in term of the perturbations, the field equations become
\begin{equation}\label{fe2}
f_0'\biggl[R^{(1)}_{\mu\nu}-\frac{R^{(1)}}{2}\eta_{\mu\nu}\biggr]-f''_{0}\biggl[R^{(1)}_{,\mu\nu}-\eta_{\mu\nu}\Box
R^{(1)}\biggr]=\frac{\mathcal{X}}{2}T^{(0)}_{\mu\nu}\,,\\
\end{equation}
where ${\displaystyle f'_0=\frac{df}{dR}\Bigl|_{R=0}}$, ${\displaystyle f''_0=\frac{d^2f}{dR^2}\Bigl|_{R=0}}$ and
$\Box={{}_{,\sigma}}^{,\sigma}$ that is  d'Alembert operator. Here, the Ricci tensor and scalar read
\begin{eqnarray}\label{approx1}
\left\{\begin{array}{ll}R^{(1)}_{\mu\nu}=h^\sigma_{(\mu,\nu)\sigma}-\frac{1}{2}\Box
h_{\mu\nu}-\frac{1}{2}h_{,\mu\nu}\\\\
R^{(1)}={h_{\sigma\tau}}^{,\sigma\tau}-\Box h \end{array}\right.
\end{eqnarray}
Now, assuming that the source is localized in a finite region, as a consequence outside this region $T_{\mu\nu}=0$ and 

\begin{eqnarray}
R^{(1)}_{\mu\nu}=\Box h_{\mu\nu}=0\,.
\label{2.17}
\end{eqnarray}

From here it is possible calculate the energy momentum tensor of gravitational field in $f(R)$-gravity, adopting the definition 
given in \cite{landau} and \cite{quadrupolo}, so that it satisfies a conservation law as required by the Bianchi identities:

 \begin{eqnarray}
t^\lambda_\alpha&=&\frac{1}{\sqrt{-g}}\left[\left(\frac{\partial\mathcal{L}}{\partial
g_{\rho\sigma,\lambda}}-\partial_\xi\frac{\partial\mathcal
{L}}{\partial
g_{\rho\sigma,\lambda\xi}}\right)g_{\rho\sigma,\alpha}+\right.\nonumber\\ &&\left.+\frac{\partial\mathcal
{L}}{\partial
g_{\rho\sigma,\lambda\xi}}g_{\rho\sigma,\xi\alpha}-\delta^\lambda_\alpha\mathcal{L}\right]\,.
\end{eqnarray}

Starting from above equation, \cite{quadrupolo} have shown that the energy momentum tensor 
consists of a sum of  a GR contribution plus a term coming from  $f(R)$-gravity

\begin{eqnarray}
t^\lambda_\alpha=f'_0{t^\lambda_\alpha}_{|_{\text{GR}}}+f''_0{t^\lambda_\alpha}_{|_{f(R)}}\,,
\end{eqnarray}

that  in term of the perturbation $h$ is 

\begin{eqnarray}
t^\lambda_\alpha&\sim&f'_0{t^\lambda_\alpha}_{|_{\text{GR}}}+f''_0\{(h^{\rho\sigma}_{\,\,\,\,\,\,\,,\rho\sigma}-\Box
h)\left [h^{\lambda\xi}_{\,
\,\,\,\,\,\,,\xi\alpha}-h^{,\lambda}_{\,\,\,\,\,\,\,\alpha}-\right.\nonumber\\&&\left.+\frac{1}{2}\delta^\lambda_\alpha(h^{\rho\sigma}_{\,\,\,\,\,\,\,,
\rho\sigma}-\Box
h)\right]-h^{\rho\sigma}_{\,\,\,\,\,\,\,,\rho\sigma\xi}h^{\lambda\xi}_{\,\,\,\,\,\,\,,\alpha}+\nonumber\\ &&+h^{\rho\sigma\,\,\,\,\,\,\,\,\,\,\lambda}_
{\,\,\,\,\,\,\,,\rho\sigma}h_{,\alpha}+h^{\lambda\xi}_{\,\,\,\,\,\,\,,\alpha}\Box
h_{,\xi}-\Box h^{,\lambda}h_{,\alpha}\}\,
\label{t}
\end{eqnarray}

To simplify the above equation the weak field limit approximation is taken into account, {\it i.e.}  the source $h_{\mu\nu}$ will be written as function of a single scalar variable $t'=t-r$, and 
as a consequence,  it will be almost plane \citep{Maggiore,quadrupolo}.

Finally, the energy momentum tensor assumes the following form
\begin{eqnarray}
t^\lambda_\alpha=\underbrace{f'_0 k^\lambda k_\alpha \left({\dot h}^{\rho\sigma}  {\dot h}_{\rho\sigma}\right)}_{GR}-\underbrace{\frac{1}{2}f''_0 \delta^{\lambda}_\alpha\left(k_\rho k_\sigma {\ddot h}^{\rho\sigma}\right)^2}_{f(R)}\,.
\label{tderi3}
\end{eqnarray}
To be more precise, the first term, depending on the choice of the constant $f'_0$, is the standard GR term,  the second 
is the $f(R)$ contribution. It is worth noticing that the order of derivative is increased of two degrees consistently to
the fact that $f(R)$-gravity is of fourth-order in the metric approach \citep{quadrupolo}.

\section{Radiated Energy}

In order to calculate the radiated energy of a gravitational waves source, 
\cite{quadrupolo} suppose that $h_{\mu\nu}$ can be represented by a discrete spectral representation.
The periodicity $T$  will be proportional to the inverse of the difference of the pair of frequency components in the wave, and 
then, the average of $\displaystyle{\frac{dE}{dt}}$ must be evaluated over an interval
equal to or greater than $T$ \citep{landau,Maggiore}. The instantaneous flux of energy through a surface of 
area $r^2 d\Omega$ in the direction ${\hat x}$ is given by
\begin{eqnarray}
\frac{dE}{dt}=r^2d\Omega {\hat x}^i t^{0i}\,,
\label{2.44}
\end{eqnarray}
and the average flux of energy can be written as
\begin{eqnarray}
\left\langle \frac{dE}{dt}\right\rangle=r^2d\Omega {\hat x}^i \langle t^{0i}\rangle\,.
\label{2.45}
\end{eqnarray}

 Defining the following moments of the mass-energy distribution:
\begin{eqnarray}
M(t)\simeq\int d^3 {\vec x}\, T^{00}(\vec{x},t)\,,
\label{2.57}
\end{eqnarray}
\begin{eqnarray}
D^{k}(t)\simeq\int d^3 {\vec x}\, x^k T^{00}(\vec{x},t)\,,
\label{2.58}
\end{eqnarray}
\begin{eqnarray}
Q^{ij}(t)\simeq\int d^3 {\vec x}\, x^i x^j T^{00}(\vec{x},t)\,,
\label{2.59}
\end{eqnarray}

and analyzing the radiation in terms of multipoles, \cite{quadrupolo} found

\begin{eqnarray}
\left\langle t^\lambda_\alpha\right\rangle&=& \left\langle f'_0 k^\lambda k_\alpha \frac{4}{r^2}\left[\left(  {\hat x}_i{\hat x}_j{ \dddot Q}^{ij}\right)^2- 2 \left( {\hat x}_k{ \dddot Q}^{ik}\right)\left( {\hat x}_j{ \dddot Q}^{ij}\right)+\right.\right.\nonumber\\ &&\left.\left.+\left({ \dddot Q}^{ij}{ \dddot Q}_{ij}\right)\right]
- f''_0 \delta^{\lambda}_\alpha \left(k_\rho k_\sigma\right)^2  \frac{2}{r^2}\left[\left(  {\hat x}_i{\hat x}_j{ \ddddot Q}^{ij}\right)^2+\right.\right.\nonumber\\&&\left.\left.- 2 \left( {\hat x}_k{ \ddddot Q}^{ik}\right)\left( {\hat x}_j{ \ddddot Q}^{ij}\right)+\left({ \ddddot Q}^{ij}{ \ddddot Q}_{ij}\right)\right]  
 \right\rangle\,.\nonumber\\
\label{2.79}
\end{eqnarray}

Using the result in eq. \eqref{2.45} and integrating over all directions they
computed the total average flux of energy due to the tensor wave,

\begin{eqnarray}
\underbrace{\left\langle \frac{dE}{dt}\right\rangle}_{(total)}&=& \frac{G}{60}\left\langle \underbrace{ f'_0  \left({ \dddot Q}^{ij}{ \dddot Q}_{ij}\right)}_{GR}
-\underbrace{f''_0 \left({ \ddddot Q}^{ij}{ \ddddot Q}_{ij}\right)}_{f(R)}
 \right\rangle\,.
 \label{2.83}\end{eqnarray}
Precisely, for  $f''_0\rightarrow 0$ and $f'_0\rightarrow \frac{4}{3}$,  eq. \eqref{2.83} becomes

\begin{eqnarray}
\underbrace{\left\langle \frac{dE}{dt}\right\rangle}_{(GR)}&=&\frac{G}{45}\left\langle { \dddot Q}^{ij}{ \dddot Q}_{ij} \right\rangle\,,
\label{2.84}\end{eqnarray}
which is the  well-known result  of GR \citep{landau,gravitation}. 
An important remark is related to the
absence of monopole and dipole terms in our considerations.
In our case, all the calculations are performed in the Jordan
frame so$ f(R)$-gravity results as a mere extension of GR
being $f(R)\,=\,R$, so any dipole
terms is null (as shown in \cite{Will} in table 10.2).
In order to put in evidence such contributions, we have
to pass in the Einstein frame where the additional degrees
of freedom of gravitational field can be recast in term of
 scalar fields. In this case, monopole and dipole terms explicitly
come out \citep{Will,jetzer,Damour}. The two approaches are
conformally equivalent but in the Einstein frame monopole and dipole terms can come out (see, e.g. \citet{PRnostro}).
%

\section{Application to pulsar binary systems}\label{sect4}
Now, our goal is to use a sample of binary pulsar systems to fix bounds on $f(R)$ parameters. To do this, we assume that the motion is Keplerian and the orbit is in 
the $(x,y)$-plane. We define $m_p$ as the pulsar mass, $m_c$ as the companion mass, and $\displaystyle{\mu=\frac{m_c m_p}{m_c+m_p}}$ as the reduced mass. 
In  $(x,y)$-plane the quadrupole matrix is

\begin{eqnarray}
{Q_{ij}} =\mu r^2 \left(
\begin{array}{cc}
\cos^2 \psi & \sin \psi \cos \psi \\
\sin \psi \cos \psi & \sin^2 \psi \\
\end{array}
\right)_{ij}\,,
\end{eqnarray}
where $i,j$ are the indexes in the orbital plane, $r$ is the  equation of the elliptic Keplerian orbit and $\psi$ is eccentric 
anomaly, and both of them are time dependent.

To compute the radiated power we need the third and fourth derivatives of quadrupole, so we must compute the time derivatives using the following
relation given in \cite{Maggiore}
\begin{eqnarray}
{\dot \psi}\,=\,\left( \frac{Gm_c}{a^3}\right)^{\frac{1}{2}}\left(1-\epsilon^2\right)^{-\frac{3}{2}}\left(1+\epsilon\cos\psi\right)^2\,,
\end{eqnarray}
where $a$ is the semi-major-axis, and $\epsilon$ is the orbital eccentricity. We obtain the following relations for the time derivatives of the quadrupole 

\begin{eqnarray}
{\dddot Q}_{11}={\cal H}_1\sin 2 \psi (\epsilon  \cos \psi+1)^2 (3 \epsilon  \cos \psi+4),
\end{eqnarray}

\begin{eqnarray}
{\dddot Q}_{22} &=& - {\cal H}_1 (8 \cos\psi + \epsilon  (3 \cos 2 \psi +5)) \times\nonumber\\&&\times \sin \psi (\epsilon  \cos \psi+1)^2,
\end{eqnarray}

\begin{eqnarray}
{\dddot Q}_{12} &=& - {\cal H}_1 (\epsilon  \cos \psi+1)^2 \times\nonumber\\&&\times (5 \epsilon  \cos \psi+3 \epsilon  \cos 3\psi+8 \cos 2 \psi),
\end{eqnarray}

\begin{eqnarray}
{\ddddot Q}_{11} &=& {\cal H}_2 \left[15\epsilon^2 \cos 4\psi+50\epsilon\cos3\psi+\right.\nonumber\\&&+\left.\left(12\epsilon^2+32\right)\cos2\psi+6\epsilon\cos\psi-3\epsilon^2\right],
\end{eqnarray}

\begin{eqnarray}
{ \ddddot Q}_{22}&=& - {\cal H}_2   \left[15\epsilon^2 \cos 4\psi+50\epsilon\cos3\psi+\right.\nonumber\\&&+\left.\left(24\epsilon^2+32\right)\cos2\psi+14\epsilon\cos\psi-7\epsilon^2\right],
\end{eqnarray}

\begin{eqnarray}
{\ddddot Q}_{12}&=&2  {\cal H}_2 \sin \psi \left[15\epsilon^2 \cos 3\psi+50\epsilon\cos2\psi+\right.\nonumber\\&&+\left.\left(33\epsilon^2+32\right)\cos\psi+30\epsilon\right],
\end{eqnarray}
where

\begin{eqnarray}
{\cal H}_1 =\frac{ (2 \pi) ^{5/3} \,G^{2/3}\, m_c\, m_p}{T^{5/3} \left(1-\epsilon ^2\right)^{5/2} \sqrt[3]{m_c+m_p}}\,\nonumber, 
\end{eqnarray}

\begin{eqnarray}
 {\cal H}_2 =  \frac{2^{2/3} \pi ^{8/3} G^{2/3} m_c\, m_p\, (\epsilon  \cos \psi+1)^3}{T^{8/3} \left(\epsilon ^2-1\right)^4 \sqrt[3]{m_c+m_p}}\,\nonumber.
\end{eqnarray}

Now, from eq. \eqref{2.83},  we can perform the time average of the radiated power writing

\begin{eqnarray}
\left\langle \frac{dE}{dt}\right\rangle=\frac{1}{T}\int^T_0 dt \frac{dE(\psi)}{dt}\,=\,\frac{1}{T}\int^{2\pi}_0 \frac{d\psi}{\dot \psi}\frac{dE(\psi)}{dt}\,,
\end{eqnarray}

and finally we get  the first time derivative of the orbital period
\begin{eqnarray}\label{periodvariation}
\dot{T}_b &=&  - \dfrac{3}{{20}}{\left( {\dfrac{T}{{2\pi }}} \right)^{ - \frac{5}{3}}}\dfrac{{\mu {G^{\frac{5}{3}}}{{({m_c} + {m_p})}^{\frac{2}{3}}}}}{{{c^5}{{(1 - {\epsilon^2})}^{\frac{7}{2}}}}} \times \nonumber\\&&
 \times \left[ {{f'_0}\left( {37{\epsilon ^4} + 292{\epsilon ^2} + 96} \right) - \dfrac{{{f''_0}{\pi ^2}{T^{ - 1}}}}{{2{{(1 + {\epsilon^2})}^3}}} \times } \right.\nonumber\\&&\left. {\times \left( {891{\epsilon ^8} + 28016{\epsilon ^6} + 82736{\epsilon ^4} + 43520{\epsilon ^2} + 3072} \right)} \right].
 \nonumber \\ 
\end{eqnarray}

In the next section we will go on to constraint the $f(R)$-theories estimating $f''_0$ from comparison between the theoretical 
predictions  of $\dot{T}_b$ and the observed one.

\subsection{Comparing theory prediction with data}

It is well known that in the relativistic binary pulsar systems there is a loss of energy due to GWs emission. 
This  energy loss, provided by GR, has been confirmed by the timing data analysis on the well known binary pulsar B1913+16 \citep{ht75a, wnt10}. 
We also must note that the systems like B1913+16 are optimal tools to constrain theories of gravitation \citep{Damour-Farese} using 
the Post-Keplerian parameters. For sake of convenience we choose the observed orbital period derivative ${\dot T_{b}}$, 
because it is one of the best observed Post-Keplerian parameters. Moreover we know, according to GR theory, 
that it is related to the foreseen orbital decay due to quadrupole gravitational radiation emitted by the binary systems.

As shown in sec. \ref{sect4}, it is possible rewrite
the first derivative of the orbital period in $f(R)$-theories of gravity. In principle, if we know exactly which Lagrangian we have to use 
to describe those type of systems, then we can predict the energy loss through GWs radiation. Here, we want to make the inverse process,
to get an estimation of the second derivative $f''_0$ imposing the strong hypothesis that the difference between the observed binary 
period variation ($ {\dot T_{b_{Obs}}} \pm \delta$) and the one obtained by the relativistic theory of gravitation, 
$\Delta{\dot T}_{b}=\dot{T}_{b_{Obs}}-{\dot T}_{GR}$,  is fully justified imposing that:  
 \begin{eqnarray}\label{f2}
 \dot{T}_{b_{Obs}}-\dot{T}_{GR}- f''_0 {\dot T}_{b_{f(R)}}=0, 
  \end{eqnarray}
 \begin{eqnarray}\label{f2bis}
  {\dot T_{b_{Obs}}}\pm\delta-{\dot T_{GR}}- f''_{0_{\pm\delta}} {\dot T_{b_{f(R)}}}=0,
 \end{eqnarray}
  and propagating  the experimental error, $\pm\delta$, on the first derivative of the observed orbital period ${\dot T_{b_{Obs}}}$,
  into an uncertainty on second derivative of gravitational theory, $f''_{0_{\pm\delta}}$. 
 What we want to emphasize is that, where GR is not able to fully explain the loss of energy by emission of GWs radiation 
 then, the additive contribution of an ETGs can provide a way to fill the gap between 
 theory and observations.
 { We also have substracted the external contributions to the orbital decay as galactic or Shklovskii acceleration when those values are available in literature}.
 Solving the eqs. \eqref{f2} and \eqref{f2bis} for $f''_0$ and $f''_{0_{\pm\delta}}$ we get an estimation of $f''_0$ and its upper and lower
 limits corresponding respectively to $\mp\delta$. In this way  $\Delta{\dot T_{b}}$   is fully explained through the orbital period 
 correction due to the ETGs  ${\dot T_{b_{f(R)}}}$. So we get:
 \begin{eqnarray}\label{f3}
 f''_0=\frac{\Delta\dot{T}_{b}}{{{\dot T}}_{b_{f(R)}}},
 \end{eqnarray}
 \begin{eqnarray}\label{f3bis}
 f''_{0_{\pm\delta}}=\frac{\Delta{\dot T}_{b_{\pm \delta}}}{{{\dot T}}_{b_{f(R)}}}, 
 \end{eqnarray}
where $\Delta{\dot T_{b_{\pm\delta}}}={\dot T_{b_{Obs}}}\pm\delta-{\dot T_{GR}}$.

Thus, among the various binary stars 
catalogues available in literature, we choose a sample of Observed Relativistic Binary Pulsars (ORBP)  such that the binary period 
$T_{b_{Obs}}$, the observed orbital period variation ${\dot T_{b_{Obs}}}$, the computed orbital period variation from general relativistic theory 
${\dot T_{GR}}$, the orbital eccentricity $\epsilon$, the masses of the components $m_{p}$ and $m_{c}$, are known with a fairly good precision.
For each system we have chosen, all previous parameters and their references are reported in Tab. \ref{table1} where we show: 
the  J-Name of the binary pulsar system, the observed orbital binary period  $T_{b_{Obs}}$ in days, 
the orbital projected semi-major axis $a sin(i) $ in light second, the orbital eccentricity $\epsilon$, the observed time variation of the orbital period 
$\dot{T}_{b_{Obs}}$, the predicted one ${\dot T}_{GR}$ (according to the GR theory), the experimental error $\pm\delta$ on 
$\dot{T}_{b_{Obs}}$ and the masses $ m_{p} $ and $  m_{c} $ of the binary system components in solar mass unit. 
         
Furthermore, in Tab.~\ref{table2} we reported: the  J-Name of the systems, the difference $ \Delta{\dot T_{GR}}$ between ${\dot T_{b_{Obs}}}$ 
and ${\dot{T}_{GR}}$ (equal to the correction   $-f''_0 {\dot T _{b_{f(R)}}}$),  the correction	$ {\dot T _{b_{f(R)}}}$,  the corresponding 
$f''_0 $ solution of \eqref{f2} shown in \eqref{f3}, the interval centered on $f''_0$ and finally, the interval centered on $f''_0 $ 
and computed from the difference: $ \frac{f''_{0_{+\delta}} - f''_{0_{-\delta}}}{2} $, where   $f''_{0_{\pm\delta}}$,  are the solutions 
of \eqref{f2} shown in \eqref{f3} taking in to account the experimental errors $\pm \delta $ on the observed orbital period variation 
${\dot T_{b_{Obs}}}$.   
       
\begin{figure*}
\centering
\subfloat[][]{
\includegraphics[width=16 cm, height=7 cm]{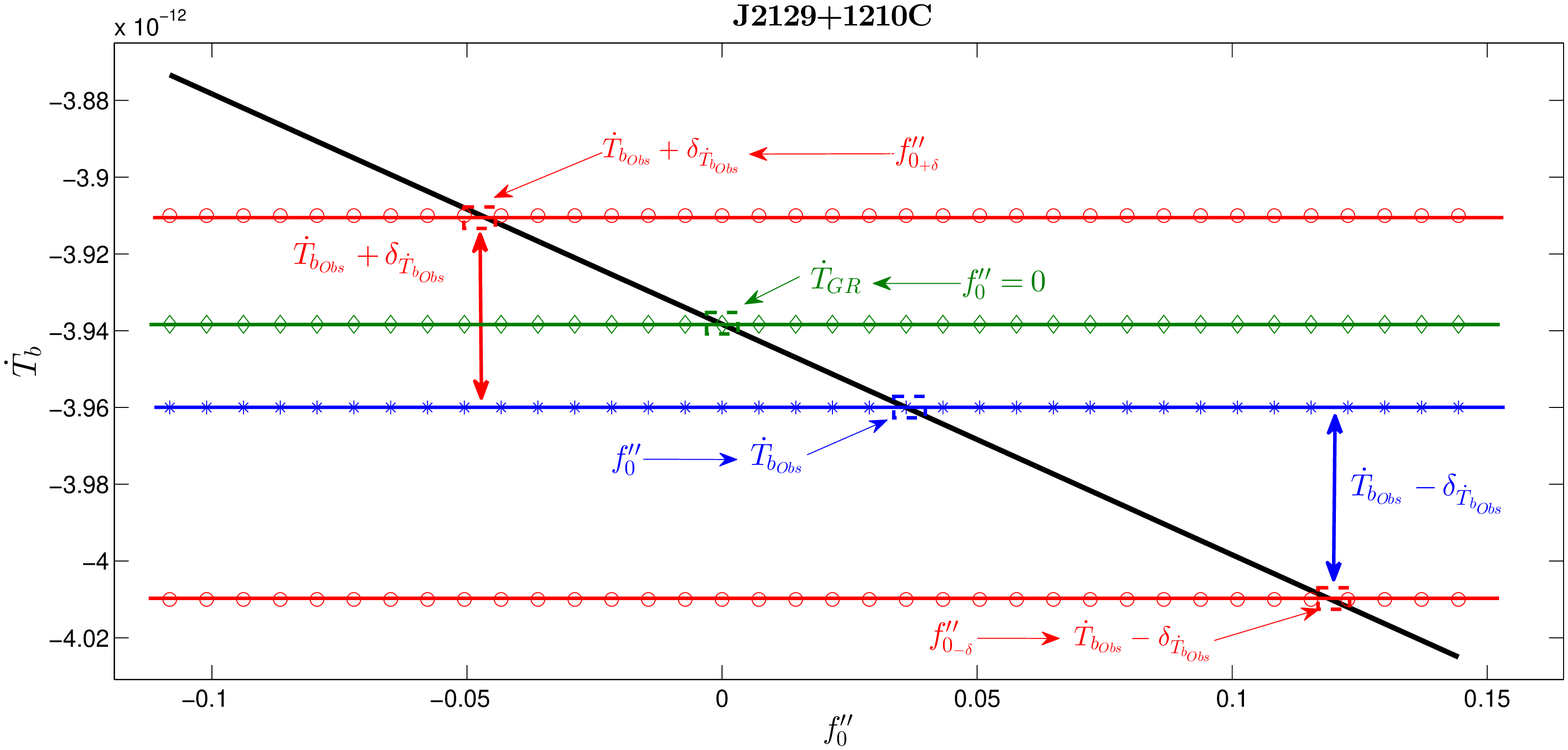}}\\[1.cm]
\subfloat[][]{
\includegraphics[width=17.3 cm, height=7 cm]{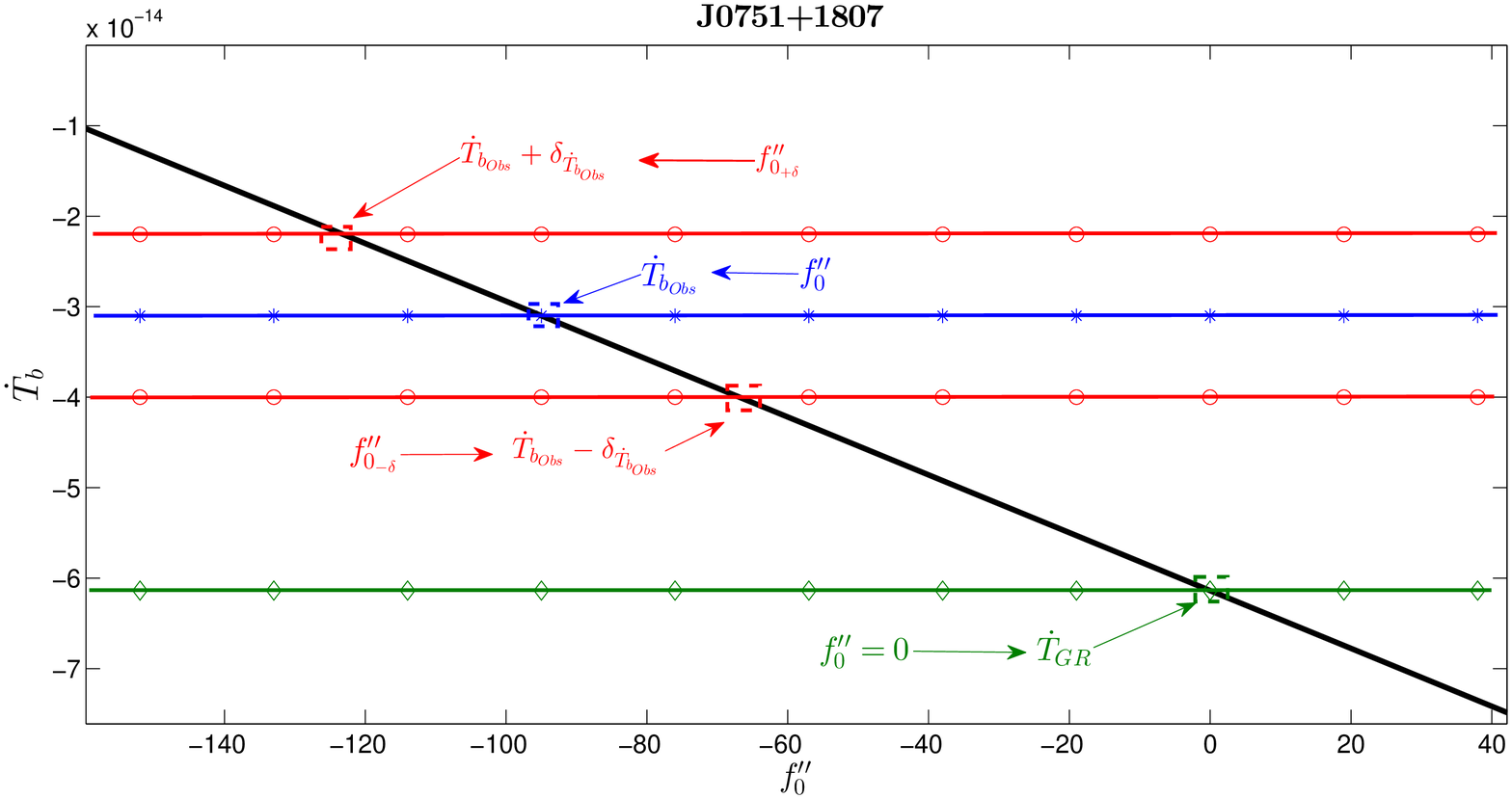}}
\caption {We report  representative results of our numerical analyses on the sample of binary pulsars we have selected.
In both figures we use the following notation: the black line shows the behavior of the first derivative of the orbital 
binary period for the $f(R)$-theory of gravitation as computed in eq. \eqref{periodvariation}; the blue line represents the observed orbital period variation ${\dot T_{b_{Obs}}}$; the red lines give the 
error band  determined by the experimental errors $\pm \epsilon$; and finally the green line is representative of the ${\dot T_{GR}}$ 
orbital period variation computed from the GR.
In the panel (a) for the system $\textbf{J2129+1210 C}$ the ${\dot T_{GR}}$ is included in the error band  determined by the 
experimental errors $\pm\delta$, so as ${\dot T_{b_{Obs}}}$. We point out the GR value of ${\dot
 T_{GR}}$ is recovered for 
$f''(r)= 0$ (green square),  while to justify the difference between ${\dot T_{b_{Obs}}}$ and
 ${\dot T_{GR}}$ we show the value 
of $f''_0$ (blue square) and its error band $f''_{0_{\pm\delta}}$ (red square) as computed in
 eqs. \eqref{f3} and \eqref{f3bis}. 
In the last panel (b) there are reported for $\textbf{J0751+1807}$ the same
 data but in this case the ${\dot T_{GR}}$ is \textbf{OUT} 
of the error band  determined by the experimental errors $\pm\delta$. It is 
 possible to see for $f''_0=0$  that  the GR value of 
${\dot T_{GR}}$ is recovered, but in this case the $f''_0$ values are much 
 greater than the previous ones.}\label{fig1}
\end{figure*}

Now, in Fig.\ref{fig1}, we report  representative results of our numerical analysis on the sample of binary pulsars we choose.
In both panels we use the following notation: the black line shows the behavior of the first derivative of the orbital 
binary period for the $f(R)$-theories of gravity as computed in eq. \eqref{periodvariation}; the blue line represents the observed orbital period variation ${\dot T_{b_{Obs}}}$; the red lines give the 
error band  determined by the experimental errors $\pm \delta$; and finally the green line is representative of the ${\dot T_{GR}}$ 
orbital period variation computed from the GR. 
For the binary pulsar system $ J2129+1210 C$, Fig.\ref{fig1}a, the orbital period variation ${\dot T_{GR}}$, computed from the 
GR, is included in experimental  error band $\pm \delta$, so as the observed orbital period variation ${\dot T_{b_{Obs}}}$.
Moreover, it is possible to see from Fig.\ref{fig1}a the GR value of ${\dot T_{GR}}$ is recovered for $f''_0= 0$ (green square), 
whilst to justify the difference $\Delta{\dot T_{GR}}$ between ${\dot T_{b_{Obs}}}$ and ${\dot T_{GR}}$ we have, from the solution of \eqref{f3},
the values shown in Fig.\ref{fig1} for $ f''_{0_{\pm\delta}}$(red square) and for $f''_0$ (blue square). 
In the panel (b) of the Fig.\ref{fig1} there is reported for $J0751+1807$ the same situation. In this 
case the ${\dot T_{GR}}$ is out of the error band  determined by the experimental errors $\pm \delta $. It is again possible to see 
for $f''_0=0$  that the GR value of ${\dot T_{GR}}$ is recovered, but in this case the $f''_0$ values are order of magnitude greater 
than the one of the well behaved case of $ J2129+1210 C $.

\begin{figure*}
\thicklines
\includegraphics[scale=0.5]{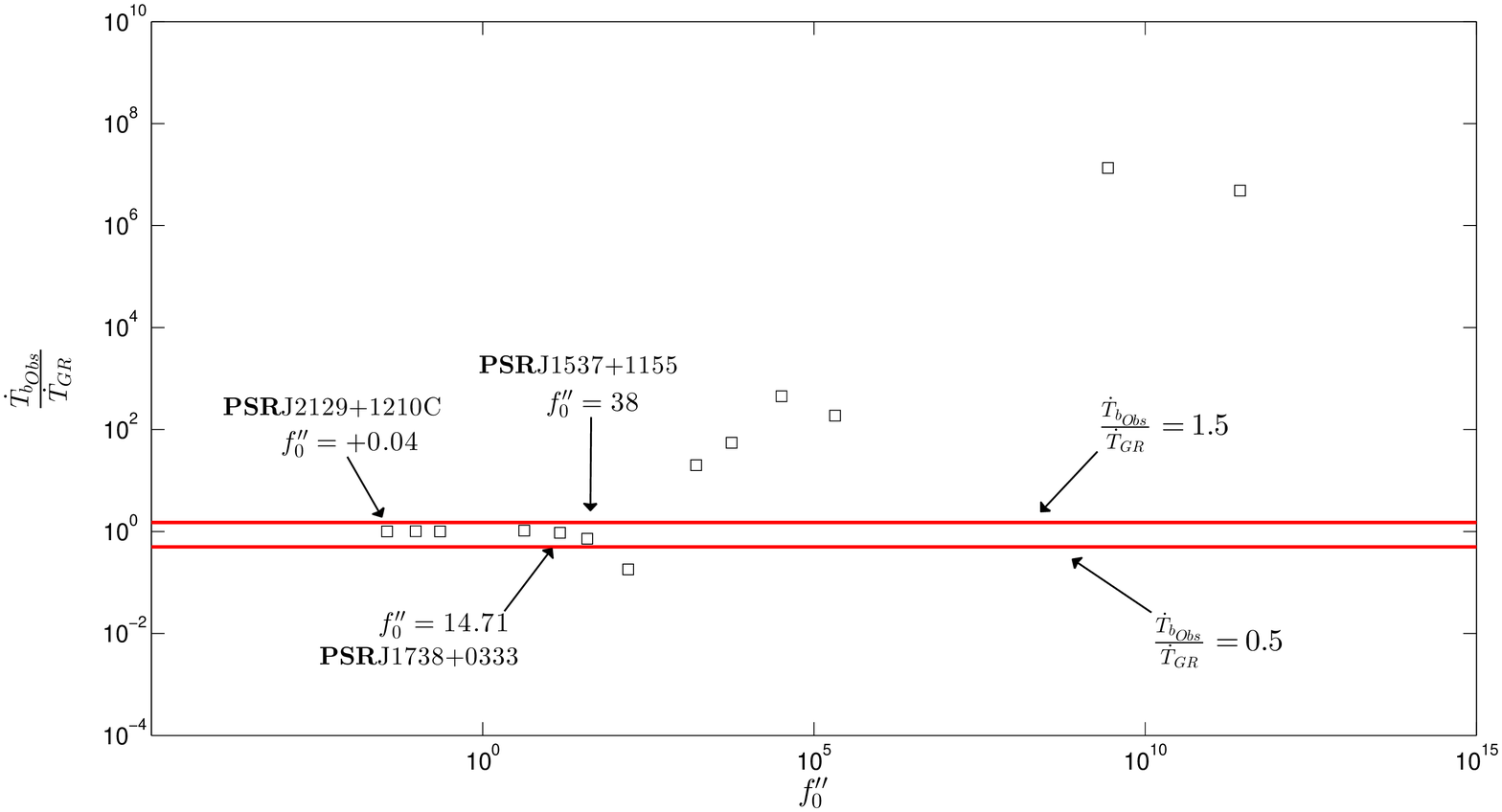}
\caption {In figure  there are shown, for sake of convenience, in logaritmic scale, the absolute  values of $f''_0$ reported in 
Tab.~\ref{table2} versus the ratio $\frac{\dot{T}_{b_{Obs}}}{\dot{T}_{GR}}$.  We must note that for five binaries the ETGs we are probing is not ruled out 
$0.04 \leq f''_0\leq \approx 38$, for those systems the difference between $\dot{T}_{GR}$ and $\dot{T}_{b_{Obs}}$ is tiny, indeed we get
$0.5 \le \frac{\dot{T}_{b_{Obs}}}{\dot{T}_{GR}} \le 1.5$. Instead for most of binaries we have $f''_0$ values that can surely rule out the theory, 
since taking account of the weak field assumption we obtain $38 \leq f''_0\leq 4 \times 10^{7}$. From this last values to the first ones, 
there is a jump of about four up to five order of magnitude on $f''_0$.}\label{fig2}
\end{figure*}

In Fig. \ref{fig2}  there are shown, for sake of convenience, in logarithmic scale,  the absolute  values of $f''_0$   reported in 
Tab.\ref{table2} versus the ratio $\frac{\dot{T}_{b_{Obs}}}{\dot{T}_{GR}}$.   
We must note that for the first six binaries in tables, the ETGs are not ruled out $0.04 \leq f''_0\leq 38$. 
For those systems we get  $0.5 \le \frac{\dot{T}_{b_{Obs}}}{\dot{T}_{GR}} \le 1.5$, the difference between $\dot{T}_{GR}$ and 
$\dot{T}_{b_{Obs}}$ can be explained adding a new contribution from the theory of gravity. Instead for most of binaries we have $f''_0$ values 
that can surely rule out the theory, since taking account of the weak field assumption we obtain $38 \leq f''_0\leq 4 \times 10^{7}$. 
From this last values to the first ones, there is a jump of about four up to five order of magnitude on $f''_0$. The origin of these strong discrepancies, 
perhaps, is due to the extreme assumption we made, to justify the difference between the observed ${\dot T_{b_{Obs}}}$ and the predicted
${\dot T_{GR}}$ using the ETGs.

\section{Discussion and Remarks}
{ We want point out in this preliminary work that, where the GR theory is not enough to explain the gap between the data and the theoretical estimation of the orbital decay, there is the possibility to extend the GR theory with a generic $f(R)$- theory to cover the gap. Here, we simply verify that this possibility exists, but there is need to compute the Post-Keplerian parameters in the $f(R)$-theory to estimate correctly the masses of the binary systems to constraint correctly the analytic parameters the ETGs.}
 In post-Minkowskian limit of analytic $f(R)$-gravity models,  the  quadrupole-radiation depends  on the masses of the two bodies, 
on the orbital parameters and on the analytic parameters of the  $f(R)$-theory as the coefficients  $f'_0$ and $f''_0$ 
of the Taylor expansion. 
A first result we present is the analytical solution of the quadrupole radiation rate in which it is possible separate
the GR contribution and the one due to the $f(R)$-gravity. We should note that the correction depends on the eccentricity
of the orbit and on the orbital period of the binary system, and specifically,  the radiation rate is a function of $f'_0$ and $f''_0$.
According to eq. \eqref{periodvariation},we have selected a sample of relativistic binary systems for which the first derivative of the orbital period is observed, we have computed the theoretical quadrupole radiation rate, and  finally we have compared it to  
binary system observations. 
  From Tab.~\ref{table2}, it is seen that the first five systems have masses determined in a manner quite reliable, while for the remaining sample,
masses are estimated by requiring that the mass of the pulsar is $1.4 M_\odot$ and, assuming for the orbital inclination one of the 
usual statistical values ($i\,=\,60^{\circ}$ or $i\,=\,90^{\circ}$ ),  and from here comes then the estimate of the mass of the companion star. So a primary cause of major 
discrepancies, not only for the ETGs, but also for the GR theory, between the variation of the observed orbital 
period  and the predicted effect of emission of gravitational waves, could be a mistake in the estimation of the masses of the system.
In addition, other causes may be attributable to the evolutionary state of the system, which, for instance, if it does not consist of two neutron stars may transfer mass from companion to the neutron star.  In our sample, there are
only five double NS that can be used to test  GR and ETGs. Taking into account of the strong hypothesis we made, the extended theory correction to 
${\dot T_{GR}} $ can also include the galactic acceleration term correction (\cite{Damour-Taylor}, \cite{Damour-Taylor1}).
Here, we give a preliminary result about the energy loss from binary systems and we show that, when the nature of the binary systems
can exclude energy losses due to trade or loss of matter, then, we can explain the gap between the first time derivative of  the observed  orbital period and the
theoretical one predicted by GR, using an analytical $f(R)$-theory of gravity.
In conclusion, to improve the estimation of the $f(R)$-coefficients, we need:
  to consider the hydrodynamic effects due to the transfer of the matter in the binary system, in order to analyze different systems from double NS; and to improve the estimations of the mass of the stars in the binary systems without prior on pulsar mass and orbital inclination.


\section*{Acknowledgments}

The authors thank S. Capozziello, A. Lauria and L. Milano for discussions and comments on the topics of this paper.
Furthermore, we warmly thank the anonymous referee for the many suggestions that made
this work more readable and formally correct.
This work has been partially funded by Ministerio de Educacion y Ciencia (Spain), grants FIS2012-30926 and N FIS2009-07238.

\begin{landscape}
\begin{table}
\caption{Data for binary Relativistic  pulsars: in the order  we reported the  J-Name of the binary pulsar system , the orbital binary period  $T_b$ in days, the orbital projected semi-major axis $a(sini)$ in light second,
 the orbital eccentricity $\epsilon$, the Observed orbital period variation
  $\dot{T}_{Obs}$, the predicted ${\dot T}_{GR}$ according to the General
   Relativity theory, the experimental error $\pm\delta$ on $\dot{T}_{Obs}$
    and the masses $ m_{p} $ and $  m_{c} $ of the binary system
    components in solar mass unit.}
\label{table1}
\begin{tabular}{|c|c|c|c|c|c|c|c|c|c|c|} 
  \hline
  \hline    
    Name	&	$ T_{b}$	&	$ a$	&	$i$ & $\epsilon$	&	${\dot T}_{b_{Obs}}$	&	${\dot
     T}_{GR}$ 	&	$\pm \delta$	&	$ m_{p} $	& $ m_{c} $ &
      $ \rm{\textbf{References}} $	\\
       &  (days) &  (lsec) &   (degrees)     &  &$(10^{-12})$ &$(10^{-12})$ & $(10^{-12})$ &$(M_{\odot})$ &$(M_{\odot}$) &  \\
    \hline
  \hline
J2129+1210C     &      0.335282049     &       2.51845         &                            &   0.681395     &   -3.96      &    -3.94      &       0.05    &   1.358    & 1.354  &    \cite{agk+90}, \cite{jcj+06} \\ 
J1915+1606	&	0.322997449	&	2.341782        &    ${54.12^{\circ}}$       &	  0.6171334	&   -2.423     &    -2.403	&	0.001	&   1.4398   & 1.3886 & \cite{ht75a},\cite{wnt10}	\\
J0737-3039A	&	0.102251562	&	1.415032        &    ${88.69^{\circ}}$       &	  0.0877775	&   -1.252     &    -1.248	&	0.017	&   1.3381   & 1.2489 &  \cite{bdp+03}, \cite{ksm+06}	\\
J1141-6545	&	0.197650959	&	1.858922        &    $73^{\circ}$            &	  0.171884	&   -0.403     &    -0.387	&	0.025	&   1.27     & 1.02   & \cite{klm+00a,bbv08}	\\
J1537+1155	&	0.420737299	&	3.7294626       &    $78.4^{\circ}$          &	  0.2736767	&   -0.138     &    -0.192	&	0.0001	&   1.3332   & 1.3452 &   \cite{stt+02,  kws03}	\\
J1738+0333      &      0.3547907399    &       0.343429        &   ${32.6^{\circ}}$         &   3.4e-7       &   -0.017     &    -0.0277    &       0.0031  &   1.46     & 0.181  &    \cite{fvf+12} \\
J0751+1807	&	0.263144267	&	0.3966127       &    $65.8^{\circ}$          &	  0.00000071	&   -0.031     &    -0.017	&	0.009	&   1.7      & 0.67  &   \cite{lzc95,nss+08} \\
J0024-7204J	&	0.120664938	&	0.0404021       &    $60^{\circ}$            &	      0	        &   -0.55      &    -0.03	&	0.13	&   1.4      & 0.024  & \cite{fck+03, clmf+00}	\\ 
J1701-3006B	&	0.144545417	&	0.2527565       &    $84.7^{\circ}$          &	      0		&   -5.12      &    -0.09	&	0.062	&   1.4      & 0.14   & \cite{pdm+03, lfrj12}	\\
J2051-0827	&	0.099110251	&	0.045052        &    $30^{\circ}$            &	      0		&   -15.5      &    -0.03	&	0.8	&   1.4      & 0.027  &   \cite{sbl+96, dlk+01}	\\
J1909-3744	&	1.533449475	&	1.8979910       &    $86.4^{\circ}$          &	  1.302E-07	&   -0.55      &    -0.003	&	0.03	&   1.57     & 0.212  &   \cite{jbv+03, vbc+09}	\\
J1518+4904	&	8.634005096	&	20.044002       &    $<47^{\circ}$           &	  0.24948451	&    0.24      &    -0.001	&	0.22	&   1.56     & 1.05   &  \cite{nst96, jsk+08}	\\
J1959+2048	&	0.381966607	&	0.0892253       &    $65^{\circ}$            &	      0		&   14.7       &    -0.003	&	0.8	&   1.4      & 0.022  &    \cite{fst88, aft94}	\\
J2145-0750	&	6.83893		&	10.164108       &                            &	  0.0000193	&    0.4       &    -0.0005	&	0.3	&   1.4      & 0.5    &    \cite{bhl+94, vbc+09}	\\
J0437-4715	&	5.74104646	&	3.36669708      &     $137.58^{\circ}$	      &	  0.00001918	&    0.159     &    -0.0004	&	0.283	&   1.76    & 0.254  &\cite{jlh+93, vbv+08} \\
J0045-7319	&	51.169451	&	174.2576        &    $44^{\circ}$            &	  0.807949	&   -3.03E+5   &    -0.02242	&	9E+3	&   1.4      & 8.8    & \cite{mmh+91, kbm+96}	\\
J2019+2425	&	76.51163479	&	38.7676297      &     $63^{\circ}$           &	  0.00011109	&  -30.0       &    -0.000006	&	60.0	&   1.33     & 0.35   &  \cite{ntf93, nss01}	\\
J1623-2631	&	191.44281	&	64.80946        &     $40^{\circ}$           &	  0.02531545	&  400.0       &    -0.000003	&	600.0	&   1.3      & 0.8    &  \cite{lbb+88, tacl99}	\\
   \hline
  \hline 
  \end{tabular}
\end{table}  
\end{landscape}

\begin{table*}
\caption{ Upper Limits of $f''_0$ correction to $ {\dot T}_{GR}$ of binary relativistic  pulsars assuming that all the loss of energy is caused by Gravitational Wave emission. We reported the  J-Name of
     the system,the	difference $ \Delta{\dot T_{GR}} $	between ${\dot T_{b_{Obs}}}$ and ${\dot
      T_{GR}}$ equal to the correction   $ -f''_0 {\dot T _{b_{f(R)}}}$  ,  the correction	$ {\dot T _{b_{f(R)}}}$ , the
       corresponding $	f''_0 $ solution of (\ref{f2}) shown in (~\ref{f3}), the interval 
       centered on $f''_0$ and computed from the difference
       $ \frac{f''_{0_{+\delta}} - 	f''_{0_{-\delta}}}{2} $	,where $f''_{0_{\pm\delta}}$
        ,	  are the solutions of (~\ref{f2}) shown in (~\ref{f3}) taking account of the
         experimental errors $\pm \delta $ on 
         the observed orbital period variation ${\dot T_{b_{Obs}}}$.}
\label{table2}
\begin{tabular}{|c|c|c|c|c|c|} 
  \hline
  \hline       
         Name	&		$ \Delta{\dot T_{GR}} $	&	$ {\dot T _{b_{f(R)}}}	$ & $  	f''_0 $	&	$   {  \pm\Delta f''_0}$	\\
  \hline
  \hline 
J2129+1210C	&	-2.17E-14	&	6.01E-13	&	 3.61E-02	&	8.32E-02	\\
J1915+1606	&	-2.04E-14	&	2.10E-13	&	 9.74E-02	&	4.77E-03	\\
J0737-3039A	&	-4.23E-15	&	1.86E-14	&	 2.28E-01	&	9.15E-02	\\
J1141-6545	&	-1.65E-14	&	3.88E-15	&	 4.25E+00	&	6.44E+00	\\
J1537+1155	&	 5.39E-14	&	1.42E-15	&	-3.79E+01	&	7.03E-02	\\
J1738+0333      &      -1.56E-15       &       1.06E-16       &       -1.47E+01       &       2.92E+01       \\ 
J0751+1807	&	 1.41E-13	&	8.98E-16	&	-15.7E+01	&	1.002E+01	\\
J0024-7204J	&	-5.22E-13	&	3.13E-16	&	 1.67E+03	&	4.15E+02	\\
J1701-3006B	&	-5.03E-12	&	8.81E-16	&	 5.71E+03	&	7.04E+01	\\
J2051-0827	&	-1.55E-11	&	4.77E-16	&	 3.24E+04	&	1.68E+03	\\
J1909-3744	&	-5.47E-13	&	2.62E-18	&	 2.09E+05	&	1.14E+04	\\
J1518+4904	&	 2.41E-13	&	3.42E-19	&	-7.05E+05	&	6.43E+03	\\
J1959+2048	&	 1.47E-11	&	1.07E-17	&	-1.38E+06	&	7.51E+04	\\
J2145-0750	&	 4.01E-13	&	1.00E-19	&	-4.00E+06	&	2.99E+06	\\
J0437-4715	&	 1.59E-13	&	1.04E-19	&	-1.57E+06	&	2.73E+06	\\
J0045-7319	&	 3.02E-07	&	1.11E-16	&	 2.74E+9	&	8.13E+07	\\
J2019+2425	&	-3.00E-11	&	1.11E-22	&	 2.71E+11	&	5.41E+11	\\
J1623-2631	&	 4.00E-10	&	2.02E-23	&	-1.98E+13	&	2.97E+13	\\
\hline
\hline
\end{tabular}
\end{table*}

\begin{thebibliography}{100}
\bibitem[\protect\citeauthoryear{Anderson et al.}{1990}]{agk+90} Anderson S.B., Gorham P.W., Kulkarni S.R. \& Prince T.A., 1990, {\it Nature}, {\bf 346}, 42-44.

\bibitem[\protect\citeauthoryear{Arzoumanian et al.}{1994}]{aft94} Arzoumanian Z., Fruchter A.S. \& Taylor J.H., 1994, {\it  ApJ},{\bf 426}, L85-L88.
 
\bibitem[\protect\citeauthoryear{Bailes et al.}{1994}]{bhl+94} Bailes M. et al., 1994,{\it  ApJ}, {\bf 425}, L41-L44.

\bibitem[\protect\citeauthoryear{Bhat et al.}{2008}]{bbv08} Bhat N.D.R., Bailes M. \& Verbiest J.P.W., 2008, {\it Phys. Rev. D}, {\bf 77(12)}, 124017.

\bibitem[\protect\citeauthoryear{Bogdanos et al.}{2010}]{greci} Bogdanos C., Capozziello S., De Laurentis M., Nesseris S., 2010, {\it Astrop. Phys. }, {\bf 34}, 236.

\bibitem[\protect\citeauthoryear{Burgay et al.}{2003}]{bdp+03} Burgay M. et al., 2003,{\it Nature}, {\bf 426}, 531-533.

\bibitem[\protect\citeauthoryear{Camilo et al.}{2000}]{clmf+00} Camilo F. et al., 2000, {\it  ApJ}, {\ bf 535},975-990

\bibitem[\protect\citeauthoryear{Capozziello and De Laurentis}{2011}]{PRnostro} Capozziello S., De Laurentis M., 2011, {\it Physics  Reports } {\bf 509}, 167.

\bibitem[\protect\citeauthoryear{Capozziello et al.}{2008}]{SCF} Capozziello S., Corda C., De Laurentis M., 2008 {\it Phys.  Lett.  B}  {\bf 669}, 255-259 .

\bibitem[\protect\citeauthoryear{Capozziello and Francaviglia}{2008}]{francaviglia} Capozziello S., Francaviglia M., 2008, {\it Gen. Rel. Grav.}   {\bf 40},357.

\bibitem[\protect\citeauthoryear{Capozziello et al.}{2009}]{faraoni} Capozziello S., De Laurentis M., Faraoni V., 2009 ,{\it The Open Astr. Jour} , {\bf 2}, 1874.

\bibitem[\protect\citeauthoryear{Damour and Esposito-Farese}{1998}] {Damour-Farese}  Damour T. and Esposito-Farese G., {\it {\it  Phys. Rev. D}}, {\bf 58}, 042001

\bibitem[\protect\citeauthoryear{Damour and Taylor}{1991}]{Damour-Taylor} Damour T., \& Taylor J.H., 1991,{\it ApJ}, {\bf 366}, 50

\bibitem[\protect\citeauthoryear{Damour and Taylor}{1992}]{Damour-Taylor1} Damour T., \& Taylor J.H.,1992, {\it  Phys. Rev. D}., {\bf 45}, 1840

\bibitem[\protect\citeauthoryear{Damour and Esposito-Farese}{1996}]{Damour} Damour \& Esposito-Farese 1996, Phys. Rev. D 54, 1474

\bibitem[\protect\citeauthoryear{De Laurentis and Capozziello}{2011}]{quadrupolo} De Laurentis M., Capozziello S., 2011, {\it Astrop. Phys. }, {\bf 35}, 257 . 

\bibitem[\protect\citeauthoryear{Doroshenko et al.}{2001}]{dlk+01} Doroshenko O.  et. al., 2001,  {\it Astron. \& Astroph}, {\bf 379}, 579-587.

\bibitem[\protect\citeauthoryear{Freire et al.}{2003}]{fck+03} Freire P.C. et al.,  2003, {\it  MNRAS}, {\bf 340}, 1359-1374.

\bibitem[\protect\citeauthoryear{Freire et al.}{2012}]{fvf+12} Freire P.C. et al.,  2012, {\it  MNRAS}, {\bf 423}, 3328.
         
\bibitem[\protect\citeauthoryear{Fruchter et al.}{1988}]{fst88} Fruchter A.S., Stinebring D.R. \& Taylor J.H., 1988, {\it Nature}, {\bf 333}, 237-239.

\bibitem[\protect\citeauthoryear{Hulse and Taylor}{1975}]{ht75a} Hulse R.A. and Taylor J.H., 1975,   {\it  ApJ}, {\bf 195}, L51-L53.
          
\bibitem[\protect\citeauthoryear{Jacoby et al.}{2003}]{jbv+03}  Jacoby B.A. et al., 2003, {\it  ApJ}, { \bf 599}, L99-L102.  
         
\bibitem[\protect\citeauthoryear{Jacoby et al.}{2006}]{jcj+06} Jacoby B.A. et al.  2006,  {\it  ApJ}, {\bf 644}, L113-L116.
         
\bibitem[\protect\citeauthoryear{Janssen et al.}{2008}]{jsk+08} Janssen G.H. et al., 2008,  {\it Astron. $\&$ Astroph}, {\bf 490}, 753-761.
 
\bibitem[\protect\citeauthoryear{Johnston et al.}{1993}]{jlh+93} Johnston S. et al., 1993, {\it Nature}, {\bf 361}, 613-615.

\bibitem[\protect\citeauthoryear{Kaspi et al.}{1996}]{kbm+96} Kaspi V.M., Bailes M., Manchester R.N., Stappers B.W. \& Bell J.F., 1996, {\it Nature}, {\bf 381}, 584-586.

\bibitem[\protect\citeauthoryear{Kaspi et al.}{2000}]{klm+00a} Kaspi V.M. et al., 2000,  {\it  ApJ}, {\bf 543}, 321-327.
         
         
\bibitem[\protect\citeauthoryear{Konacki et al.}{2003}]{kws03} Konacki M., Wolszczan A. \& Stairs, I. H., 2003 , {\it  ApJ}, {\bf 589}, 495-502.

\bibitem[\protect\citeauthoryear{Kramer et al.}{2006}]{ksm+06} Kramer M. et al., 2006, {\it Science}, {\bf 314}, 97-102.

         
\bibitem[\protect\citeauthoryear{Landau and Lifshitz}{1962}]{landau} Landau L.D., Lifshitz E.M., 1962, {\it The Classical Theory of Fields},  Addison-Wesley Pub.Co., Inc., Reading.

\bibitem[\protect\citeauthoryear{Lundgren et al.}{1995}]{lzc95} Lundgren S.C., Zepka A.F. \& Cordes J.M., 1995, {\it  ApJ}, {\bf 453}, 419-423.

\bibitem[\protect\citeauthoryear{Lynch et al.}{2012}]{lfrj12} Lynch R.S., Freire P.C.C., Ransom S.M. \& Jacoby B.A., 2012, {\it  ApJ}, {\bf 745}, 109.

\bibitem[\protect\citeauthoryear{Lyne}{2004}]{lbk+04} Lyne A.G., 2004, {\it  Science}, {\bf 303}, 1153-1157.
         
\bibitem[\protect\citeauthoryear{Lyne et al.}{1988}]{lbb+88} Lyne A.G., et al., 1988, {\it Nature}, {\bf 332}, 45-47.
         


\bibitem[\protect\citeauthoryear{Maggiore}{2007}]{Maggiore} Maggiore M., 2007, {\it Gravitational Wawes: Theory and Experiments},  Oxford Univ. Press, Oxford.

\bibitem[\protect\citeauthoryear{ Manchester et al.}{1991}]{mlr+91} Manchester R.N. et al. 1991, {\it Nature}, {\bf 352}, 219-221.

\bibitem[\protect\citeauthoryear{McConnell et al.}{1991}]{mmh+91} McConnell D. et al., 1991, {\it  MNRAS}, {\bf 249}, 654-657.

\bibitem[\protect\citeauthoryear{Naef and Jetzer}{2011}]{jetzer} Naef \& Jetzer 2011, Phys. Rev. D 84, 024027.
         
\bibitem[\protect\citeauthoryear{Nice et al.}{2008}]{nss+08} Nice D.J. et al., 2008. , {\it  AIPCP}, {\bf 983}, 453-458.
         
\bibitem[\protect\citeauthoryear{Nice et al.}{1993}] {ntf93} Nice D.J., Taylor J.H. \& Fruchter, A. S., 1993, {\it  ApJ}, {\bf 402}, L49-L52.
         
\bibitem[\protect\citeauthoryear{Nice et al.}{2001}]{nss01} Nice D.J., Splaver E.M. \& Stairs, I. H., 2001, {\it  ApJ}, {\bf 549}, 516-521.

\bibitem[\protect\citeauthoryear{Nice et al.}{1996}]{nst96}Nice D.J., Sayer R.W. \& Taylor, J. H., 1996, {\it  ApJ}, {\bf 466}, L87-L90.

\bibitem[\protect\citeauthoryear{Nojiri and Odintsov}{2011}]{PRsergei} Nojiri S., Odintsov S.D., 2011, {\it Physics Reports} {\bf 505}, 59.

\bibitem[\protect\citeauthoryear{Nojiri and Odintsov}{2007}]{nojiodi} Nojiri S., Odintsov S.D., 2007 {\it Int. J. Geom. Meth. Mod. Phys.} {\bf 4}, 115.

\bibitem[\protect\citeauthoryear{Possenti et al.}{2003}]{pdm+03} Possenti A. et al., 2003, {\it  ApJ}, {\bf 599}, 475-484. 

\bibitem[\protect\citeauthoryear{Stairs et al.}{2002}]{stt+02} Stairs, I. H., Thorsett, S. E., Taylor, J. H.,  Wolszczan, A. 2002, {\it  ApJ},{\bf 581}, 501

\bibitem[\protect\citeauthoryear{Stappers et al.}{1996}]{sbl+96} Stappers B.W. et al., 1996, {\it  ApJ}, {\bf 465}, L119-L122.

\bibitem[\protect\citeauthoryear{Thorsett et al.}{1999}]{tacl99} Thorsett S.E., Arzoumanian, Z., Camilo, F. \& Lyne, A. G., 1999, {\it  ApJ}, {\bf 523}, 763-770.
         
\bibitem[\protect\citeauthoryear{van Kerkwijk et al.}{2011}]{kbk+11} van Kerkwijk M.H., Breton R.P. and  Kulkarni S.R., 2011, {\it  ApJ}, {\bf 728}, 95.
         
\bibitem[\protect\citeauthoryear{Verbiest  et al.}{2008}]{vbv+08} Verbiest J.P.W. et al., 2008,  {\it  ApJ}, {\bf 679}, 675-680.
         
\bibitem[\protect\citeauthoryear{Verbiest  et al.}{2009}]{vbc+09} Verbiest J.P.W.  et al., 2009,  {\it  MNRAS}, {\bf 400}, 951-968.

\bibitem[\protect\citeauthoryear{Weinberg}{1972}]{gravitation} Weinberg S., 1972 ,{\it "Gravitation and Cosmology"}, John Wiley \& Sons, Inc., New York.

\bibitem[\protect\citeauthoryear{Weisberg et al.}{2010}]{wnt10}  Weisberg J.M., Nice D.J. and Taylor J.H., 2010, {\it  ApJ}, {\bf 722},1030-1034.

\bibitem[\protect\citeauthoryear{Will}{1993}]{Will} Will C.M, 1993, "{\it Theory and experiment in gravitational physics}" Cambridge University Press, UK.

\bibitem[\protect\citeauthoryear{Wolszczan}{1990}] {wol90a} Wolszczan, A.,1990, {\it IAU Circ.} {\bf No. 5073}.
\end{thebibliography}
\end{document}